\documentclass{article}
\usepackage{spconf,amsmath,graphicx}
\usepackage{booktabs}
\usepackage{multirow}
\usepackage{hyperref}

\title{Enhancing Open-Set Speaker Identification through Rapid Tuning with Speaker Reciprocal Points and Negative Sample}
\name{Zhiyong Chen, Zhiqi Ai, Xinnuo Li and Shugong Xu}
\address{School of Communication and Information Engineering, Shanghai University}
\begin{document}
%
\maketitle
\begin{abstract}
This paper introduces a novel framework for open-set speaker identification in household environments, playing a crucial role in facilitating seamless human-computer interactions. Addressing the limitations of current speaker models and classification approaches, our work integrates an pretrained WavLM frontend with a few-shot rapid tuning neural network (NN) backend for enrollment, employing task-optimized Speaker Reciprocal Points Learning (SRPL) to enhance discrimination across multiple target speakers. Furthermore, we propose an enhanced version of SRPL (SRPL+), which incorporates negative sample learning with both speech-synthesized and real negative samples to significantly improve open-set SID accuracy. Our approach is thoroughly evaluated across various multi-language text-dependent speaker recognition datasets, demonstrating its effectiveness in achieving high usability for complex household multi-speaker recognition scenarios. The proposed system enhanced open-set performance by up to 27\% over the directly use of efficient WavLM base+ model. For detailed information on open-sourced implementation in our project website\footnote{\url{https://github.com/zhiyongchenGREAT/speaker-reciprocal-points-learning}}.
\end{abstract}
\section{Introduction}
In household environments, the use of AI agents to accurately identify speakers through speech, a method not limited by posture or obstacles, is increasing. This is crucial for enabling seamless interaction between humans and computers, including interactions with large language model (LLM). Speaker identification, a subtask of speaker recognition, employs learning approaches to precisely recognize in-house speakers by effectively utilizing the voice print characteristics of target speakers\cite{bai2021speaker}.

Recent advancements in speaker recognition have featured the development of advanced speaker models, such as TDNN-based \cite{sigona2024validation}, ResNet-based \cite{peng2023attention}, and state-of-the-art self-supervised learning models like WavLM \cite{chen2022wavlm, peng2023attention} and other audio pretrained front-end models (LMs) \cite{jung2024espnet}. Despite their general applicability in representing speakers, these models primarily focus on binary classification \cite{han2023exploring}, leaving room for improvement in speaker identification (SID) algorithms for high usability in complex household multi-speaker scenarios \cite{ohi2021deep}.



In SID, it is assumed that test utterances originate from a set of pre-enrolled speakers, employing strategies like multi-class classification loss \cite{li2020speaker, hong2020combining}, prototype learning loss for few-shot learning \cite{chen2021adversarial, li2023few}, or graph-based learning method \cite{tong2022graph}. However, these approaches predominantly target \textit{closed-set} classification, which may not adequately serve real-world applications facing the challenge of \textit{open-set} speaker identification, where the goal is to optimize detection of target-set speakers with contrastive accuracy and robustness to outliers.



Recent advancements in open-set SID have explored using multiple Probabilistic Linear Discriminant Analysis (PLDA) modules for outlier detection \cite{wilkinghoff2020open} and prototype-based loss \cite{kishan2022openfeat} for improved identification. However, these methods, not inherently designed for open-set recognition, often fail to achieve optimal performance or lack the streamlined characteristic of neural network (NN)-based approaches \cite{kishan2022openfeat}.

In response to the increasing interest in open-set learning capabilities across various fields of pattern recognition \cite{chen2021adversarial}, we propose a novel framework that integrates an effcient pretrained audio front-end model (WavLM base+) with a few-shot rapid tuning NN backend for enrollment. This approach employs task-optimized Speaker Reciprocal Points Learning (SRPL) to improve open-set discrimination across multiple target speakers. Furthermore, we introduce an enhanced version of SRPL (SRPL+), incorporating negative sample learning with both speech-synthesized and real negative samples to significantly improve the accuracy of open-set SID. Our main contributions include:

\begin{itemize}
\item Introduction of SRPL for open-set speaker identification, facilitating rapid tuning enrollment on pretrained audio front-end model (WavLM base+).
\item Incorporation of negative sample learning with SRPL+, utilizing speech-synthesized and real negative samples for enhanced identification accuracy.
\item Comprehensive evaluations on various multi-language text-dependent datasets, both qualitatively and quantitatively.
\end{itemize}


\section{Speaker Reciprocal Points Learning for Open-set Speaker Identification}

\begin{figure}[t]
  \centering
  \includegraphics[width=0.9\linewidth]{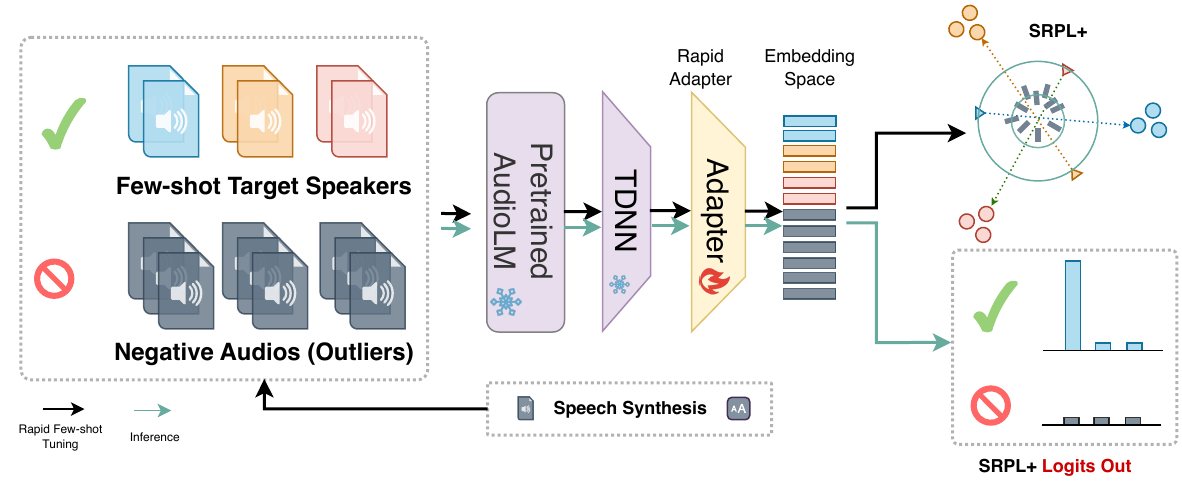}
  \caption{Illustration of open-set speaker identification architecture: customization via audio large model (LM) with SRPL-based backend rapid tuning.}
  \label{fig:sid}
\end{figure}

To enhance speaker distinction in specific domains, we propose the use of lightweight neural adaptation models. These models refine speaker embeddings for improved alignment with target scenarios. Figure \ref{fig:sid} illustrates our approach, which combines a WavLM frontend \cite{chen2022wavlm} and a TDNN model pretrained on speaker recognition tasks. The process generates an initial embedding $\mathbf{Emb_{LM}}$, which the Lightweight Adapter transforms into a domain and speaker-specific embedding $\mathbf{Emb_{s}}$:

\begin{equation}
\begin{aligned}
& \mathbf{Emb_{s}} = Adapter(TDNN(\mathbf{Emb_{LM}}))
\end{aligned}
\end{equation}

This efficient adaptation serves as the enrollment phase in speaker recognition.

\subsection{Rapid Downstream Tuning Approach with Speaker Reciprocal Points Learning (SRPL)}
\label{sec:SRPL}
\begin{figure}[hb]
  \centering  \includegraphics[width=\linewidth]{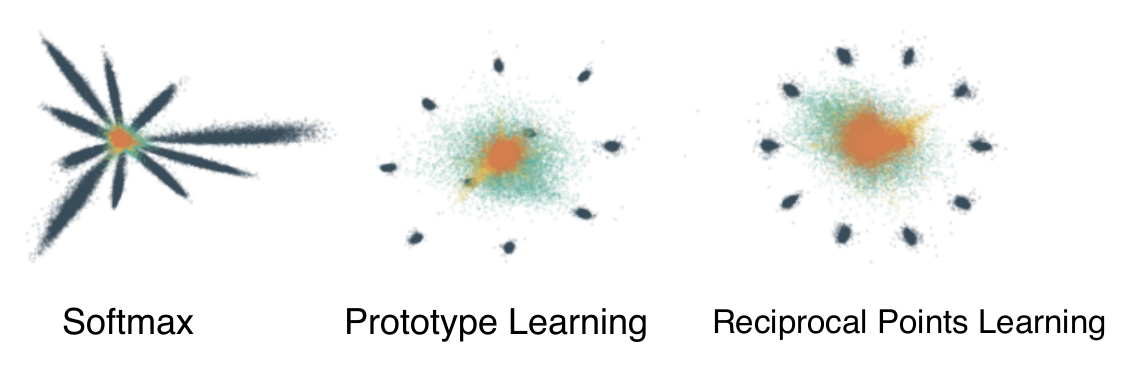}
  \caption{Conceptual illustration of the embedding space for various open-set training losses. Figure is adapted from \cite{chen2021adversarial}.}
  \label{fig:Learning}
\end{figure}

Our approach to open-set SID leverages advanced learning strategies, emphasizing few-shot learning for rapid tuning to enhance speaker recognition with limited data. Moving beyond traditional prototype learning loss methods \cite{li2023few, kishan2022openfeat} and drawing inspiration from reciprocal points learning theory \cite{chen2021adversarial}, we adapt reciprocal points learning for speaker recognition, termed Speaker Reciprocal Points Learning (SRPL). SRPL effectively distinguishes both known and unknown speakers, ensuring an optimal distribution of known speaker embeddings and establishing a dedicated area for characterizing unknown speakers, as shown in Figure \ref{fig:Learning}. Based on generalized pretrained WavLM base+ models, SRPL enhances the system's robustness for open-set recognition.

\begin{figure}[t]
  \centering
  \includegraphics[width=0.9\linewidth]{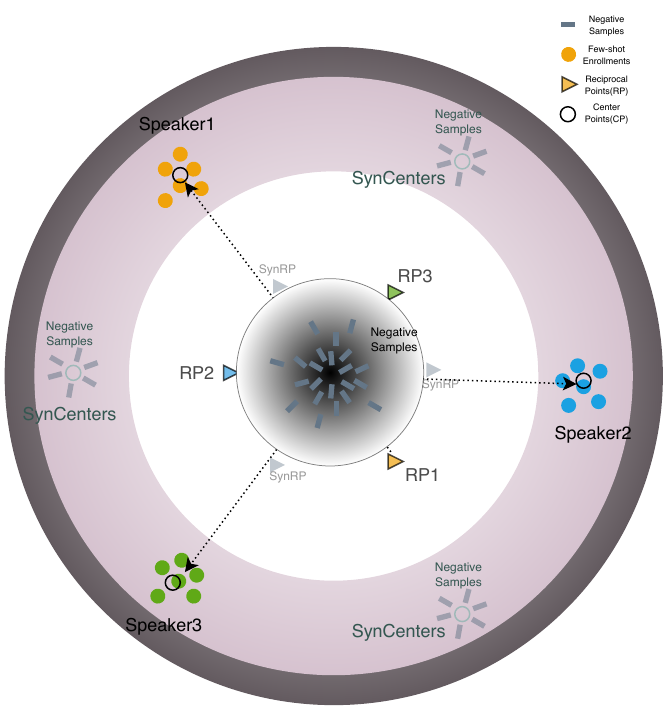}
    \caption{SRPL and its enhanced SRPL+ with integration of negative samples.}
  \label{fig:sid_learning}
\end{figure}

As illustrate in Figure \ref{fig:sid_learning}, SRPL aims to maximize the distance between learnable embeddings and Reciprocal Points (RPs), with task-specific optimizations tailored for speaker recognition. Since the audio front-end model utilizes angle-based optimization, diverging from the original approach, we calculate the distance between RPs and speaker embeddings using the \textit{inner product}, with the goal of maximizing this distance. Consequently, the probability of identifying the target speaker is evaluated based on the distance between the adapted speaker embeddings and RPs.
\begin{equation}
p(k | \mathbf{Emb_{s}}) = \frac{e^{ - \mathbf{Emb_{s}} \cdot \mathbf{RP}^k}}{\sum_{i=1}^{K} e^{ -\mathbf{Emb_{s}} \cdot \mathbf{RP}^i}},
\label{eq:probsid}
\end{equation}
\begin{equation}
\mathcal{L}_s(\mathbf{Emb_{s}}, y=k; \theta) = -\log p(k | \mathbf{Emb_{s}}),
\end{equation}
where $\mathbf{RP}^{k}$ are the RPs for the corresponding known speakers $k$, where $k$ is the speaker class.

The core of SRPL focuses on optimizing the speaker manifold by confining or reserving a general unknown pool $\mathbf{U}$ within a predefined radius $R$, by referencing to RPs. The method aims to ensure that the maximum distance from the unknown sample set $\mathbf{U}$ to all reciprocal points in RP-set $\Re$, does not exceed $R$.

\begin{equation}
\max(Dist(\mathbf{U}, \mathbf{\Re})) \leq R.
\end{equation}

Omitting detailed deductions and proofs as outlined in reciprocal points learning theory \cite{chen2021adversarial}, we can safely formulate the well-established and feasible equivalent training loss as a marginal-based loss.

\begin{equation}
\begin{aligned}
& D_e(\mathbf{Emb_{s}}, \mathbf{RP}^{k}) = \parallel\mathbf{\mathbf{Emb_{s}}} - \mathbf{RP}^{k}\parallel_{2}^{2}, \\
& \mathcal{L}_r(\mathbf{Emb_{s}}; \theta, y=k) = \max(D_e(\mathbf{Emb_{s}}, \mathbf{RP}^{k}) - R, 0),\\
\end{aligned}
\end{equation}
where we utilize Euclidean distance $D_e(\cdot,\cdot)$ to constrain the magnitude of unknowns. This optimization does not rely on prior knowledge of unknown negative data. Interestingly, the goal can be achieved by optimizing known speaker embeddings in relation to their reciprocal points. The reciprocal points loss therefore represent by $\mathcal{L}_r$.

To enhance learning process further, we further introduce center-focused learning to improve the alignment of target speaker data with its clustering center points $\mathbf{CP}^k$, a well-established method for enhancing training stability \cite{bai2021speaker}:

\begin{equation}
\mathcal{L}_c(\mathbf{Emb_{s}}, y=k; \theta) = -\log \frac{e^{  \mathbf{Emb_{s}} \cdot \mathbf{CP}^k}}{\sum_{i=1}^{K} e^{ \mathbf{Emb_{s}} \cdot \mathbf{CP}^i}},
\end{equation}

This is achieved by calculating the center (mean) points of each speaker's embeddings. The center focusing loss is represent by $\mathcal{L}_c$. Putting them together, the SRPL learning algorithm and SRPL loss $\mathcal{L}_{SRPL}$ is performed as:
\begin{equation}
\mathcal{L}_{SRPL}(\mathbf{Emb_{s}}, y; \theta) = \mathcal{L}_s(\cdot) + \lambda_r \mathcal{L}_r(\cdot) + \lambda_c \mathcal{L}_c(\cdot).
\label{eq:total_o}
\end{equation}
Detailed implementation code of this method can be found on the project website$^1$.

\subsection{SRPL Enhancement with Negative Samples (SRPL+)}\label{sec:enhac}
\subsubsection{Zero-shot Speech Synthesis Process for Negative Samples Synthesis}
\label{sec:vc}
For typical household few-shot SID tasks, acquiring a pool of unknowns $\mathbf{U}$ is often not feasible, leading to SRPL optimization primarily with known samples. While this approach is effective, SRPL inherently can support enhancement through the integration of negative samples that represent unknowns, further improving its performance. As suggested by \cite{chen2021adversarial}, employing a GAN for negative sample generation is a recognized method. However, our evaluation finds such approach is resource-intensive and unstable. To streamline the process for speaker identification, we propose a zero-shot text-to-speech (ZS-TTS) synthesis module to generate multi-speaker identity speech samples, as illustrate in Figure \ref{fig:sid}.

Negative samples are generated using randomly selected reference speakers from the LibriTTS dataset \cite{qin2023openvoice}, ensuring their voiceprints are distinct from those used in evaluations, thus forming $\mathbf{U}_{syn}$. Synthesis speed controls are applied to enrich the diversity of negative examples. The synthesis content are the target keywords in the text-dependent evaluation. This approach enhances the SRPL framework, termed SRPL+. We utilize the VITS baed ZS-TTS system TSCM-VITS \cite{chen2024optimizing}, a zero-shot TTS model trained on the LibriTTS corpus, facilitating efficient and diverse negative sample creation.



\subsubsection{Learning Enhancement with Negative Audio Instance}\label{sec:ad_samp}
For SRPL+, we refine our approach to utilize audio-level negative samples within the SRPL learning framework, as illustrated in Figure \ref{fig:sid_learning}. This integration comprises two main processes. First, it employs negative samples to enhance performance by generating synthetic speakers, increasing the number of synthesized reciprocal points (SynRPs) and as a supplement to calculate more center points for synthesized speakers (SynCPs). This is proved to effectively improved the robustness of the learning. Next, negative data primarily supports the adversarial learning process, using synthetic text-dependent vast speaker negative samples $u_{syn} \in \mathbf{U}_{syn}$ to maximize their distance from all RPs, effectively targeting maximum entropy $H(\cdot)$. This strategy realizes the SRPL+ optimization goal, enhancing the model's ability to distinguish between known and unknown speakers, and the enhanced SRPL loss $\mathcal{L}_{SRPL+}$ defined as:


\begin{equation}
\mathcal{L}_{SRPL+}(\mathbf{Emb_{s}},y,\mathbf{U}_{syn}; \theta) =  \mathcal{L}_{SRPL}(\cdot) - \lambda_{ns} H(u_{syn}, \Re).
\end{equation}

\section{Experiments Settings}
\label{sec:exp}
This section details our experimental setup, datasets, metrics, and procedures, offering an overview necessary for replication. Full details are on our project website$^1$.

\subsection{Datasets}
\begin{itemize}
\item \textbf{Qualcomm Speech Dataset:} This dataset contains speech utterances of four English keywords by 50 speakers, designed for evaluating our few-shot open-set SID adaptation/enrollment approach in a household multi-class classification scenario \cite{lee2023phonmatchnet}. We apply 5-fold cross-validation, calculating the mean performance across datasets. For each split, a selection of 10 speakers is reserved for testing targets, and 15 for testing outliers, with the remainder preserved (which is used as \textit{real negative samples} for SRPL+ testing). No overlap between outlier testing data and any training-stage speaker data is ensured. Therefore all speakers are iteratively tested as target or outlier in the evaluation stage.
\item \textbf{FFSVC HiMia:} Featuring utterances of the Chinese wake-up command 'Hi Mia' by over 80 speakers \cite{qin2022far}, this dataset's dev-set captures a range of distances and microphone-array recordings. We focus on the near-field portion of the FFSVC dev-set for our experiments. Similar to the Qualcomm setup, 10 speakers are reserved for target testing, and 15 for testing outliers, employing 5-fold cross-validation. The other experimental setup is the same as that used with the Qualcomm Speech Dataset.


\end{itemize}
\subsection{Metrics}
For evaluating our open-set SID tasks, we focus on multi-target classification metrics common to the pattern recognition and speech processing domain. Area Under the Receiver Operating Characteristic Curve (AUROC or AUC) is used as a threshold-independent metric for SID \cite{nassif2021casa}. It measures the probability that a model correctly ranks a randomly chosen target example higher than a randomly chosen negative example by plotting the true positive rate against the false positive rate.

Nonetheless, AUC may not fully capture the unique challenges of open-set recognition, which requires effectively rejecting outliers while ensuring target speakers are accurately identified. To address this, we employ the Open Set Classification Rate (OSCR) \cite{chen2021adversarial}, a more robust metric that simulates open-set recognition dynamics:

\begin{equation}
\resizebox{1.05\linewidth}{!}{
$CCR(TH) = \frac{|\{x \in TestData^{k} \mid \arg\max_{k} P(k|x) = k \cap P(k|x) \geq TH\}|}{|TestData^{k}|},$
}
\end{equation}
\begin{equation}
FPR(TH) = \frac{|\{x \mid x \in Unknown \cap \max_k P(k|x) \geq TH\}|}{|Unknown|},
\end{equation}
where $TH$ is a threshold. OSCR calculates the area under the curve mapping the Correct Classification Rate (CCR) for known classes to the False Positive Rate (FPR) for unknown data, offering a threshold-independent evaluation for open-set.
\subsection{Training and Inference Details}
For our open-set SID system, we employ rapid fine-tuning with SRPL or SRPL+ in few-shot settings. During the enrollment phase, each target speaker contributing 20 enrolled utterances alongside 20 synthesized utterances for model rapid tuning. For SRPL+, we incorporate 1000+ negative samples from unknown speakers, synthesized via ZS-TTS (SynNeg) or sourced from the unused dataset portions (RealNeg). The system adapts using a 3-layer MLP, linearly transforming to K-way speaker outputs, and applies the SRPL/SRPL+ loss. Fine-tuning the adapter employs an SGD optimizer for 100 epochs, a process completed in mere seconds. Rather than using cosine similarity, we evaluate test speaker embeddings against reciprocal points using the calculated logits from Equation (\ref{eq:probsid}). All reciprocal points (RPs) and center points (CPs) are learnable and hyperparameter $\lambda$s are all set equally to 1, without careful search to achieve effectiveness.


\begin{table*}[t]
\caption{Experimental results of open-set SID tasks.}
\centering
\resizebox{0.8\linewidth}{!}{
\begin{tabular}{@{}llccclccc@{}}
\toprule
\multicolumn{1}{c}{\multirow{2}{*}{Method}} &  & \multicolumn{2}{c}{Open-set Eval} & Close-set Eval &  & \multicolumn{2}{c}{Open-set Eval} & Close-set Eval \\ \cmidrule(lr){3-4} \cmidrule(lr){5-4} \cmidrule(lr){7-8} \cmidrule(lr){9-8}
\multicolumn{1}{c}{} &  & AUC(\%)↑ & OSCR(\%)↑ & ACC(\%)↑ &  & AUC(\%)↑ & OSCR(\%)↑ & ACC(\%)↑ \\ \midrule
\multicolumn{1}{c}{Qualcomm Speech \cite{lee2023phonmatchnet}} &  & \multicolumn{3}{c}{5Way} &  & \multicolumn{3}{c}{10Way} \\ \midrule
WavLM-TDNN CosineDirect \cite{chen2022wavlm} &  & 84.09 & 83.61 & 99.54 &  & 83.81 & 81.49 & 96.06 \\ \midrule
\quad SoftmaxTune \cite{hong2020combining} &  & 71.33 & 70.95 & 99.07 &  & 66.46 & 66.91 & 98.91 \\
\quad ProtoTypeTune \cite{li2023few} &  & 71.54 & 71.39 & 99.54 &  & 88.87 & 88.45 & 98.91 \\
\quad OpenFEAT \cite{kishan2022openfeat} &  & 73.23 & 72.88 & 99.07 &  & 88.05 & 87.35 & 98.24 \\
\quad SRPL (Ours) &  & 84.12 & 84.95 & 99.53 &  & 89.53 & 89.16 & 98.47 \\ \midrule
\quad SRPL+(SynNeg) &  & 92.51 & 92.48 & \underline{99.90} &  & 92.06 & 91.25 & \underline{98.69} \\
\quad SRPL+(RealNeg) &  & \textbf{95.73} & \textbf{94.25} & \underline{99.21} &  & \textbf{95.40} & \textbf{94.26} & \underline{98.81} \\ 
\midrule \midrule
\multicolumn{1}{c}{FFSVC HiMia \cite{qin2022far}} &  & \multicolumn{3}{c}{5Way} &  & \multicolumn{3}{c}{10Way} \\
\midrule
WavLM-TDNN CosineDirect \cite{chen2022wavlm} &  & 85.07 & 80.27 & 93.50 &  & 78.61 & 67.24 & 84.25 \\ \midrule
\quad SoftmaxTune \cite{hong2020combining} &  & 75.87 & 75.86 & 99.80 &  & 88.66 & 87.60 & 97.25 \\
\quad ProtoTypeTune \cite{li2023few} &  & 75.77 & 74.67 & 97.50 &  & 88.87 & 86.90 & 95.25 \\
\quad OpenFEAT \cite{kishan2022openfeat} &  & 80.10 & 78.98 & 97.00 &  & 85.63 & 82.83 & 94.75 \\
\quad SRPL(Ours) &  & 89.12 & 88.91 & 99.00 &  &  91.28 & 90.27 & 98.50 \\ \midrule
\quad SRPL+(SynNeg) &  & \textbf{95.71} & \textbf{94.97} & \underline{99.00} &  & \textbf{95.48} & \textbf{94.83} & \underline{99.00} \\
\quad SRPL+(RealNeg) &  & \underline{93.04} & \underline{91.93} & \underline{98.00} &  & \underline{94.32} & \underline{92.71} & \underline{97.80} \\ \bottomrule
\end{tabular}
}
\label{tab:SID_m}
\end{table*}

\section{Results}
\subsection{Comparative Evaluation of SRPL with Baselines}

Table \ref{tab:SID_m} shows our SRPL system's performance relative to various baseline methods. A basic comparison point is the WavLM TDNN system, which applies cosine similarity for speaker identification, using mean embeddings of enrollment audios and cosine scores for probability output across different channels. This approach reduces SID to a binary classification problem without optimizing for multi-class scenarios. It does not reasonably detects close-set scenarios, particularly with up to 10 speakers in the FFSVC evaluation, and struggles with outliers, showing its vulnerability in open-set contexts.

More advanced methodologies, such as Softmax fine-tuning \cite{hong2020combining}, prototype learning \cite{li2023few}, and OpenFEAT learning \cite{kishan2022openfeat}—which builds on prototype learning by incorporating negative samples—are specifically designed for rapid tuning and enrollment in speaker identification. Our SRPL system surpasses these methods, including the WavLM-TDNN baseline, in both Prototype and OpenFEAT approaches. Its superiority is evident not just in close-set evaluations, where many methods excel in targeting speaker accuracy, but crucially in open-set metrics, underscoring SRPL's robust performance in 5-way and 10-way classification tasks.

With the introduction of negative samples, SRPL+ significantly outperforms baseline methods in OSCR scores, highlighting its advancement. This enhancement is also observed with synthetic negative samples (SRPL+(SynNeg)), which demonstrates excellent results. Moreover, in close-set metrics, SRPL+ maintains strong performance, showcasing its comprehensive effectiveness in open-set speaker identification.
\subsection{Supplementary Analyses for SRPL for Open-set SID}

\begin{figure}[ht!]
  \centering
  \includegraphics[width=\linewidth]{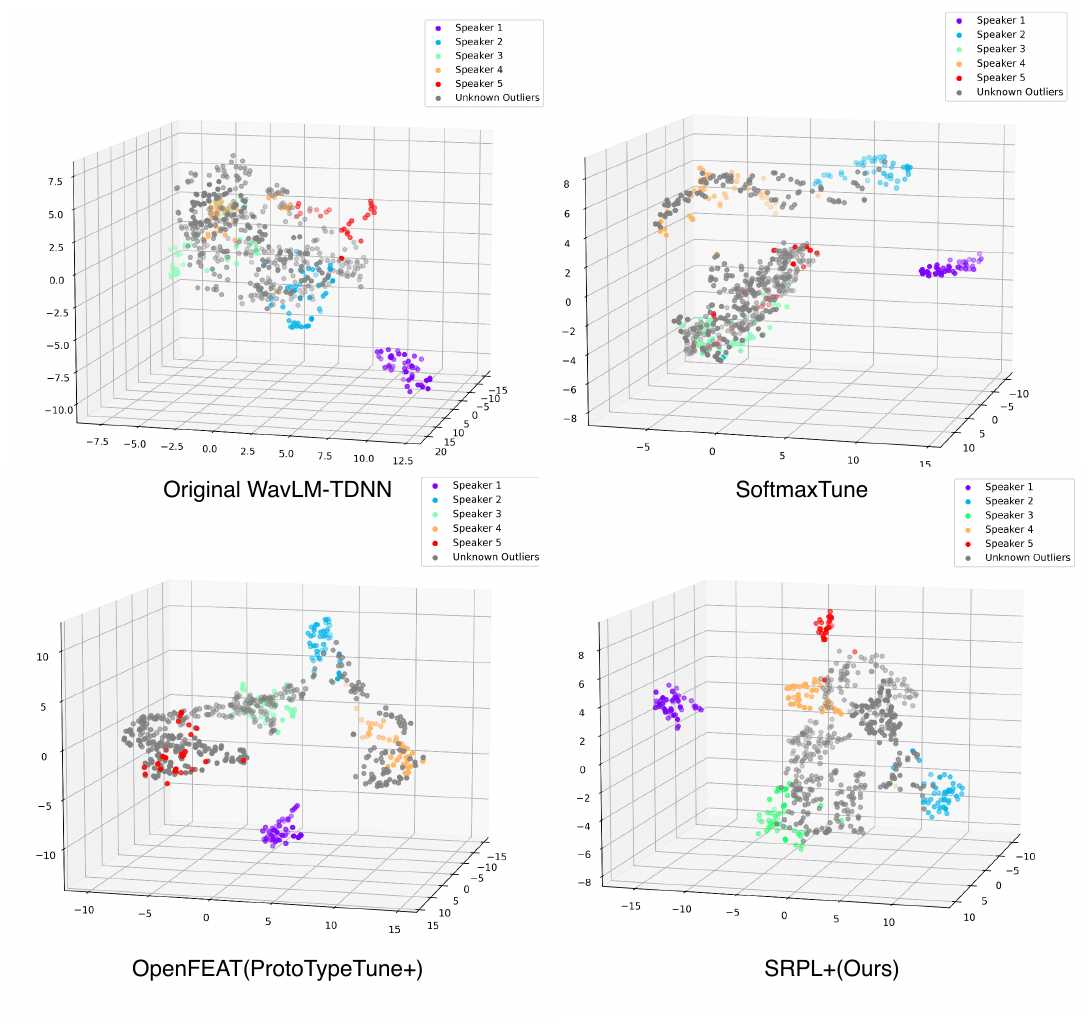}
  \caption{t-SNE visualization of speaker embeddings for targets and outliers in testing datasets for SRPL+ and baseline systems.}
  \label{fig:sid_arpemb}
\end{figure}

Figure \ref{fig:sid_arpemb} showcases a t-SNE visualization of speaker embeddings, revealing that, for the WavLM-TDNN baseline, embeddings are not distinctively separated from unknowns, indicating a lack of discriminability. In contrast, SRPL's rapid tuning markedly enhances the discrimination of speaker embeddings, effectively clustering unknowns into specific regions and improving the grouping of known target speakers. The softmax optimization, focusing on the angle of distribution, performs suboptimally compared to SRPL. While the OpenFEAT method, an advancement on prototype learning, shows improved class cohesion when learning with negatives, it falls short in separating negative instances.

Table \ref{tab:spk_abl} evaluates the efficacy of SRPL's components. Utilizing an angular-aware distance metric, concentrating on center points, and incorporating negative samples for both synthesized centers and adversarial learning significantly bolster SRPL's overall performance in speaker recognition. The backend tuning process is efficient, requiring only 200 seconds to safely converge on GPU. The use of 1000+ negative data for optimal performance adds only an additional 80 seconds, highlighting its practical feasibility.


\begin{table}[ht!]
\caption{Ablation Studies and Training Cost Analysis.}
\centering
\resizebox{\linewidth}{!}{
\begin{tabular}{@{}lccc@{}}
\toprule
\multicolumn{1}{c}{\multirow{2}{*}{Method}} & \multicolumn{2}{c}{Open-set (10-Way)} &  Tuning Cost \\ \cmidrule(l){2-3}
\multicolumn{1}{c}{} & AUC(\%)↑ & OSCR(\%)↑ & (GPU) \\ \midrule
\textbf{SRPL+} & \textbf{95.40} & \textbf{94.26} & \multirow{2}{*}{280s} \\
\quad w/o SynCenters & 93.31 & 92.83&  \\
\midrule
\textbf{SRPL}  & \underline{89.53} & \underline{89.16}& \multirow{3}{*}{200s} \\
\quad w/o CenterFocus & 88.05 & 88.30&  \\
\quad \quad w/o SpkTaskOptimize & 87.50 & 86.01&  \\
\bottomrule
\end{tabular}
}

\label{tab:spk_abl}
\end{table}

\section{Conclusions}
The challenges of current speaker recognition systems for open-set speaker identification are evident. This study introduces Speaker Reciprocal Points Learning (SRPL), a novel algorithm specifically optimized for open-set SID, leveraging the state-of-the-art pretrained front-end model combined with a few-shot rapid tuning neural network backend for enrollment. This approach models unknown speakers using known speaker samples. It is further enhanced by SRPL+, effectively trained with negative samples, including zero-shot TTS-synthesized samples or real samples collected. Our method demonstrates a significant performance improvement over existing techniques, notably enhancing usability in household environments. Moving forward, we aim to refine our methodology to more datasets and researching more fine-tuning methods to improve its performance, and incorporating multimodal audio-visual approaches or other methods to enhance system robustness.



\bibliographystyle{IEEEbib}
\bibliography{mybib}

\begin{thebibliography}{10}

\bibitem{bai2021speaker}
Zhongxin Bai and Xiao-Lei Zhang,
\newblock ``Speaker recognition based on deep learning: An overview,''
\newblock {\em Neural Networks}, vol. 140, pp. 65--99, 2021.

\bibitem{sigona2024validation}
Francesco Sigona and Mirko Grimaldi,
\newblock ``Validation of an ecapa-tdnn system for forensic automatic speaker recognition under case work conditions,''
\newblock {\em Speech Communication}, p. 103045, 2024.

\bibitem{peng2023attention}
Junyi Peng, Old{\v{r}}ich Plchot, Themos Stafylakis, Ladislav Mo{\v{s}}ner, Luk{\'a}{\v{s}} Burget, and Jan {\v{C}}ernock{\`y},
\newblock ``An attention-based backend allowing efficient fine-tuning of transformer models for speaker verification,''
\newblock in {\em 2022 IEEE Spoken Language Technology Workshop (SLT)}. IEEE, 2023, pp. 555--562.

\bibitem{chen2022wavlm}
Sanyuan Chen, Chengyi Wang, Zhengyang Chen, Yu~Wu, Shujie Liu, Zhuo Chen, Jinyu Li, Naoyuki Kanda, Takuya Yoshioka, Xiong Xiao, et~al.,
\newblock ``Wavlm: Large-scale self-supervised pre-training for full stack speech processing,''
\newblock {\em IEEE Journal of Selected Topics in Signal Processing}, vol. 16, no. 6, pp. 1505--1518, 2022.

\bibitem{jung2024espnet}
Jee-weon Jung, Wangyou Zhang, Jiatong Shi, Zakaria Aldeneh, Takuya Higuchi, Barry-John Theobald, Ahmed~Hussen Abdelaziz, and Shinji Watanabe,
\newblock ``Espnet-spk: full pipeline speaker embedding toolkit with reproducible recipes, self-supervised front-ends, and off-the-shelf models,''
\newblock {\em arXiv preprint arXiv:2401.17230}, 2024.

\bibitem{han2023exploring}
Bing Han, Zhengyang Chen, and Yanmin Qian,
\newblock ``Exploring binary classification loss for speaker verification,''
\newblock in {\em ICASSP 2023-2023 IEEE International Conference on Acoustics, Speech and Signal Processing (ICASSP)}. IEEE, 2023, pp. 1--5.

\bibitem{ohi2021deep}
Abu~Quwsar Ohi, Muhammad~F Mridha, Md~Abdul Hamid, and Muhammad~Mostafa Monowar,
\newblock ``Deep speaker recognition: Process, progress, and challenges,''
\newblock {\em IEEE Access}, vol. 9, pp. 89619--89643, 2021.

\bibitem{li2020speaker}
Ruirui Li, Jyun-Yu Jiang, Carrie Wu, Chu-Cheng Hsieh, and Andreas Stolcke,
\newblock ``Speaker identification for household scenarios with self-attention and adversarial training,''
\newblock in {\em Interspeech 2020}, 2020.

\bibitem{hong2020combining}
Qian-Bei Hong, Chung-Hsien Wu, Hsin-Min Wang, and Chien-Lin Huang,
\newblock ``Combining deep embeddings of acoustic and articulatory features for speaker identification,''
\newblock in {\em ICASSP 2020-2020 IEEE International Conference on Acoustics, Speech and Signal Processing (ICASSP)}. IEEE, 2020, pp. 7589--7593.

\bibitem{chen2021adversarial}
Guangyao Chen, Peixi Peng, Xiangqian Wang, and Yonghong Tian,
\newblock ``Adversarial reciprocal points learning for open set recognition,''
\newblock {\em IEEE Transactions on Pattern Analysis and Machine Intelligence}, vol. 44, no. 11, pp. 8065--8081, 2021.

\bibitem{li2023few}
Yanxiong Li, Hao Chen, Wenchang Cao, Qisheng Huang, and Qianhua He,
\newblock ``Few-shot speaker identification using lightweight prototypical network with feature grouping and interaction,''
\newblock {\em IEEE Transactions on Multimedia}, 2023.

\bibitem{tong2022graph}
Fuchuan Tong, Siqi Zheng, Min Zhang, Yafeng Chen, Hongbin Suo, Qingyang Hong, and Lin Li,
\newblock ``Graph convolutional network based semi-supervised learning on multi-speaker meeting data,''
\newblock in {\em ICASSP 2022-2022 IEEE International Conference on Acoustics, Speech and Signal Processing (ICASSP)}. IEEE, 2022, pp. 6622--6626.

\bibitem{wilkinghoff2020open}
Kevin Wilkinghoff,
\newblock ``On open-set speaker identification with i-vectors.,''
\newblock in {\em Odyssey}, 2020, pp. 408--414.

\bibitem{kishan2022openfeat}
KC~Kishan, Zhenning Tan, Long Chen, Minho Jin, Eunjung Han, Andreas Stolcke, and Chul Lee,
\newblock ``Openfeat: Improving speaker identification by open-set few-shot embedding adaptation with transformer,''
\newblock in {\em ICASSP 2022-2022 IEEE International Conference on Acoustics, Speech and Signal Processing (ICASSP)}. IEEE, 2022, pp. 7062--7066.

\bibitem{qin2023openvoice}
Zengyi Qin, Wenliang Zhao, Xumin Yu, and Xin Sun,
\newblock ``Openvoice: Versatile instant voice cloning,''
\newblock {\em arXiv preprint arXiv:2312.01479}, 2023.

\bibitem{chen2024optimizing}
Zhiyong Chen, Zhiqi Ai, Youxuan Ma, Xinnuo Li, and Shugong Xu,
\newblock ``Optimizing feature fusion for improved zero-shot adaptation in text-to-speech synthesis,''
\newblock {\em EURASIP Journal on Audio, Speech, and Music Processing}, vol. 2024, no. 1, pp. 28, 2024.

\bibitem{lee2023phonmatchnet}
Yong-Hyeok Lee and Namhyun Cho,
\newblock ``Phonmatchnet: phoneme-guided zero-shot keyword spotting for user-defined keywords,''
\newblock {\em arXiv preprint arXiv:2308.16511}, 2023.

\bibitem{qin2022far}
Xiaoyi Qin, Ming Li, Hui Bu, Shrikanth Narayanan, and Haizhou Li,
\newblock ``Far-field speaker verification challenge (ffsvc) 2022: Challenge evaluation plan,'' 2022.

\bibitem{nassif2021casa}
Ali~Bou Nassif, Ismail Shahin, Shibani Hamsa, Nawel Nemmour, and Keikichi Hirose,
\newblock ``Casa-based speaker identification using cascaded gmm-cnn classifier in noisy and emotional talking conditions,''
\newblock {\em Applied Soft Computing}, vol. 103, pp. 107141, 2021.

\end{thebibliography}

\end{document}